**Modelling the Establishment of PAR Protein Polarity in the One-Cell *C. elegans* Embryo**

Filipe Tostevin[1#] & Martin Howard[2]*

[1] Department of Mathematics, Imperial College London, London SW7 2AZ, UK
[2] Department of Computational and Systems Biology, John Innes Centre, Norwich NR4 7UH, UK

* Corresponding author. Email: martin.howard@bbsrc.ac.uk

[#] Current address: FOM Institute for Atomic and Molecular Physics, 1098 SJ Amsterdam, The Netherlands

**ABSTRACT**

At the one-cell stage, the *C. elegans* embryo becomes polarized along the anterior-posterior axis. The PAR proteins form complementary anterior and posterior domains in a dynamic process driven by cytoskeletal rearrangement. Initially, the PAR proteins are uniformly distributed throughout the embryo. Following a cue from fertilization, cortical actomyosin contracts towards the anterior pole. PAR-3/PAR-6/PKC-3 (the anterior PAR proteins) become restricted to the anterior cortex. PAR-1 and PAR-2 (the posterior PAR proteins) become enriched in the posterior cortical region. We present a mathematical model of this polarity establishment process, in which we take a novel approach to combine reaction-diffusion dynamics of the PAR proteins coupled to a simple model of actomyosin contraction. We show that known interactions between the PAR proteins are sufficient to explain many aspects of the observed cortical PAR dynamics in both wild-type and mutant embryos. However, cytoplasmic PAR protein polarity, which is vital for generating daughter cells with distinct molecular components, cannot be properly explained within such a framework. We therefore consider additional mechanisms that can reproduce the proper cytoplasmic polarity. In particular we predict that cytoskeletal asymmetry in the cytoplasm, in addition to the cortical actomyosin asymmetry, is a critical determinant of PAR protein localization.

**INTRODUCTION**

During the one-cell stage, the *C. elegans* embryo becomes highly polarized along the anterior-posterior axis. This polarization restricts cytoplasmic P-granules and other germline-determining factors to the posterior daughter cell (1,2), identifying it as the germline precursor (3). The first mitotic division of the embryo is asymmetric and correct polarization is required to displace the division plane from mid-cell towards the posterior pole. Screens for defects in early division and polarization (see, for example, (4)) have identified many proteins involved in polarity in the *C. elegans* embryo, including the PAR proteins (PAR-1 through PAR-6) and also PKC-3, an atypical protein kinase C. The anterior and posterior regions of the cell are marked by the accumulation of PAR-3, PAR-6 and PKC-3 at the anterior cortex and PAR-1 and PAR-2 at the posterior cortex (5). The PAR proteins and their interactions are highly conserved, and regulate cell polarity in many different organisms and cell types (reviewed by (6-8)).

Recently it was found that the establishment of polarity is a highly dynamic process (9-11). The experimentally observed dynamics are summarized in Fig. 1*A*. Before the onset of polarity, the PAR proteins are uniformly distributed throughout the cell and can be detected in both the cortex and cytoplasm (9). Fertilization by the sperm near the posterior pole of the embryo triggers the establishment of polarity (12), causing contraction of a network of cortical actomyosin towards the anterior pole (11). This reorganization of actomyosin is accompanied by a gradual restriction of the anterior PAR proteins to the anterior half of the embryo (9,11). The posterior PAR proteins occupy the newly vacated posterior cortex (9,10). Competitive antagonistic interactions between the two groups of PAR proteins may help to maintain the segregated anterior and posterior domains (9,13). The polarity defects observed in *par* mutants are in part due to the disruption of the polarity establishment process and actomyosin contraction (10,11).

The considerable complexity of these dynamics calls for a mathematical description of the system that can quantitatively investigate possible mechanisms of polarization. While the PAR proteins have been extensively studied experimentally in different organisms, mathematical modelling of these systems has not previously been undertaken. In this paper we construct such a model for polarity establishment in the one-cell *C. elegans* embryo. The PAR protein interactions and random diffusive motion can be readily described by a system of nonlinear reaction-diffusion equations. However, the distributions of the PAR proteins may also be influenced by the dynamics of the cortical actomyosin network, with the anterior PAR proteins becoming restricted to the contractile anterior cortical region. The dynamics and regulation of the actomyosin network is highly complex, potentially involving actin polymerization, myosin motor activity, cross-linking proteins and interaction with the cell membrane. Many of these effects and interactions are not well understood. The available evidence also suggests that the anterior PAR proteins enhance actomyosin contractility (11), although the mechanism by which this takes place is not known. Our aim in modelling the actomyosin dynamics is to capture the effects on the PAR distributions at a similar spatial scale as our reaction-diffusion dynamics, without making detailed assumptions about specific interactions. We therefore construct a highly simplified model of actomyosin contraction that reproduces the experimental results on cellular length scales, while neglecting smaller-scale details that do not significantly affect the global protein distributions. We couple this description to our reaction-diffusion model, thereby enabling us to model the feedback between contractile actomyosin and the PAR protein distributions.

Initially, we develop a simple model that includes only the previously reported interactions between the PAR proteins together with diffusion and actomyosin contraction. We find that these interactions allow us to reproduce many features of the PAR system that are observed in vivo, including the polar cortical domains and the cortical dynamics in *par* mutant phenotypes. However, this model is unable to correctly reproduce the polarized distributions of the PAR proteins in the cytoplasm and the resulting polarity of cytoplasmic components such as MEX-5/6 (9), which are vital for the different development of the two daughter cells. We conclude that the observed cytoplasmic polarity is not simply a consequence of polarization of the cortex. Instead some additional, as yet unknown, mechanism is required in order to ensure appropriate cytoplasmic polarity.

We therefore consider ways in which the basic model can be modified to better capture this effect. We show that it is unlikely that cortical and cytoplasmic flows or protein degradation play a significant role in determining the observed distributions. Instead, we propose that cytoskeletal asymmetry in the cytoplasm, as well as on the cortex, is responsible for generating the appropriate polarity by sequestering the PAR proteins in the appropriate part of the cytoplasm. This mechanism is in accord with the known experimental data and is able to reproduce the observed dynamics in both wild-type and *par* mutant embryos. Our modelling highlights the establishment of cytoplasmic polarity as an area where our current understanding of the PAR system is incomplete. Finally, we propose experiments that could test our predicted cytoplasmic immobilization mechanism.

**RESULTS**

**Reaction-diffusion model of known interactions**

We first construct a mathematical model of the previously reported interactions between the PAR proteins. To simplify our model somewhat we separate the PAR proteins into anterior and posterior groups, as PAR protein types within each group are normally colocalized within the embryo (14-16). The variable *A* will represent the densities of the anterior PAR proteins PAR-3, PAR-6 and PKC-3, that have been suggested to form a complex (15,16). We will let *P* represent the densities of the posterior PAR proteins PAR-1 and PAR-2, although it is not known whether PAR-1 and PAR-2 interact directly. The PAR proteins can be cortically-localized ($A_m$, $P_m$) or in the cytoplasm ($A_c$, $P_c$). Reactions between proteins within each group tend to promote association - all of the anterior proteins are required for the cortical localization of PAR-6 and PKC-3 (9,15,16), and PAR-2 may enhance the cortical localization of PAR-1 (14). Interactions between the two groups tend to be antagonistic, and mutual negative feedback from the localization of each group onto the other has been proposed to explain in part the segregation of the PAR proteins into anterior and posterior domains (9). The shared properties of association/antagonism by members of each group make it advantageous to model the PAR system at the level of the anterior/posterior protein groups, rather than modelling each protein type separately. A model of the latter type would be significantly more complex, but with little additional predictive advantage.

Crucial to the polarity establishment process is rearrangement and contraction of the cortical actomyosin network towards the anterior pole (11). The density of this contractile actomyosin domain is represented in our model by *a*. Levels of actomyosin that remain at the posterior cortex are much lower than those at the anterior (11,17) and cortical ruffling is eliminated at the posterior, suggesting that the observed global contraction is largely driven by the anterior domain. Consequently, we do not include this posterior actin domain in the model. Since the embryo is polarized only along the anterior-posterior axis, we restrict the model to one dimension.

Both the anterior and posterior PAR proteins dynamically associate with the cortex (10). We will assume that this cortical dynamics is the result of both diffusion of cortical proteins and exchange of proteins between the cortex and cytoplasm. We further assume that the anterior PAR proteins associate at an increased rate with the contractile actomyosin region. This is consistent with the observation that during polarity establishment in posterior *par* mutants, the anterior PAR proteins remain restricted to the anterior cortex (9). This association may be due to the presence of CDC-42, which is required for maximal cortical localization of the anterior PAR proteins (18,19), or some other difference between the cortical actomyosin in the anterior and posterior domains. The anterior PAR proteins may not associate directly with the actomyosin cytoskeleton itself, since the myosin and anterior PAR localization patterns are slightly different (11). In addition to spontaneous dissociation, PKC-3 phosphorylates PAR-2 (13) and we assume this promotes removal of the posterior PAR proteins from the cortex. We also allow PAR-1 to stimulate dissociation of the anterior PAR proteins from the cortex, possibly through phosphorylation of PAR-3. Evidence for this reaction has been found in PAR homologues in other species (20), and a similar process has been proposed to occur in *C. elegans* (13). In this way, cortical localization of one group acts to exclude the other, and hence provides an effective positive feedback to its own accumulation. The cortical exclusion reactions likely require the 14-3-3 protein PAR-5 (9). We do not model PAR-5 explicitly since it is uniformly localized throughout the cortex and cytoplasm (21). We also do not include PAR-4, since its interactions with other PAR proteins and its effect on their distributions is not known.

Fig. 1*B* summarizes the interaction network. Our model consists of reaction-diffusion equations for the PAR protein interactions. The PAR proteins are also coupled to a simple model of cortical actomyosin contraction by incorporating enhanced cortical binding of the anterior PAR proteins in the presence of contractile actomyosin. The resulting equations are

$$\frac{\partial A_m}{\partial t} = D_m \frac{\partial^2 A_m}{\partial x^2} + (c_{A1} + c_{A2}a)A_c - c_{A3}A_m - c_{A4}A_m P_m \qquad (1a)$$

$$\frac{\partial A_c}{\partial t} = D_c \frac{\partial^2 A_c}{\partial x^2} - (c_{A1} + c_{A2}a)A_c + c_{A3}A_m + c_{A4}A_m P_m \qquad (1b)$$

$$\frac{\partial P_m}{\partial t} = D_m \frac{\partial^2 P_m}{\partial x^2} + c_{P1}P_c - c_{P3}P_m - c_{P4}A_m P_m \qquad (1c)$$

$$\frac{\partial P_c}{\partial t} = D_c \frac{\partial^2 P_c}{\partial x^2} - c_{P1}P_c + c_{P3}P_m + c_{P4}A_mP_m \ . \tag{1d}$$

The first term on the right hand side of Eqs. 1a-d represents undirected protein diffusion. The remaining terms describe the various reactions in the model. $(c_{A1}+c_{A2}a)A_c$ represents cortical association of the anterior PAR proteins, which is enhanced in the presence of contractile actomyosin. The density of actomyosin, $a$, is calculated from our actomyosin model, as described in the next section. Similarly, $P_c$ associates with the cortex through the $c_{P1}P_c$ term. $c_{A3}A_m$ and $c_{P3}P_m$ give spontaneous dissociation of the anterior and posterior PAR proteins. The terms $c_{A4}A_mP_m$ and $c_{P4}A_mP_m$ represent competitive exclusion of the cortical $A$ and $P$ groups. Since these binding and dissociation terms represent exchange between the cytoplasm and cortex, they appear in the equations for both cortical and cytoplasmic densities with opposite signs. Note that the above model does not incorporate production or degradation of the PAR proteins.

**Modelling actomyosin contraction**

In the model described above, actomyosin dynamics feeds back onto the PAR distributions through the varying density of contractile actomyosin. As the anterior actomyosin network contracts its density increases, leading to enhanced binding of the anterior PAR proteins. In order to quantify this effect, we now need to construct a simplified model of the actomyosin activity. Such a model will enable us to calculate the density of actomyosin in the contractile region, while neglecting detailed actomyosin dynamics which do not affect the PAR distributions on a cellular scale. We emphasize that the polarization of the actomyosin cytoskeleton is crucial in our model in order to break the symmetry of the system. If the actomyosin dynamics are removed, no spatial variation in the PAR protein densities can develop.

We assume that the actomyosin network is initially under tension. A polarization cue from the sperm (12,22) is believed to cause a down-regulation of the actomyosin network near the posterior pole. While it is possible that the polarity signal also affects the PAR proteins directly, this effect is not necessary in our model for polarity establishment. Once the symmetry of the network has been broken in this way, the remaining network is unstable and contracts towards the anterior. We therefore choose to model the effective dynamics of the actomyosin network as an elastic medium. The convergent flows of myosin observed in kymographs are consistent with such a global contraction model (11). To introduce positive feedback from the anterior PAR proteins onto contractility (11), we will allow the elastic properties of the system to vary depending on $A_m$. We simplify the elastic model further by assuming that, rather than $A_m$ altering the local elastic properties, the properties of the actomyosin network as a whole depend only on the *total* amount of $A_m$ in the contractile region. This assumption also implies that the actomyosin network contracts uniformly. This is a reasonable assumption, since, in our simulations, the density of $A_m$ in the anterior contractile domain is relatively constant, varying by only up to 20% from the average in this region. However, in reality, actomyosin contraction is non-uniform on short length scales, giving rise to dynamic features such as cortical ruffling and pseudocleavage. Nevertheless, we find that our coarse-grained model gives good agreement with measurements of the cortical dynamics over cellular length scales. The assumption of homogeneity also makes the model much simpler to analyse and allows us to easily compute the contraction dynamics. Relaxing this assumption would require significantly more complex model while not giving qualitatively different behaviour at a cellular scale.

The resulting dynamical equations are simply those of a uniform spring. In the subcellular environment viscous forces dominate over inertial forces. The motion of the spring will therefore be overdamped, and we neglect the second-order term in the equation of motion. In this limit of large damping, the dynamics of the spring are determined by four physical quantities: the Young's modulus, $E$, which is the ratio of the applied stress to the resulting strain; the cross-sectional area, $\tilde{A}$; the damping coefficient, $\gamma$, which determines the rate of energy dissipation; and the natural length, $\lambda$, the length of the spring when no force is applied. Assuming that $\tilde{A}$ and $\gamma$ are constant as the spring expands and contracts the length of the spring, $l(t)$, will be given by

$$\frac{dl}{dt} = v_l(t) = -\frac{\varepsilon}{\lambda(t)}\left(l(t) - \lambda(t)\right), \qquad (2)$$

where $\varepsilon = E\tilde{A}/\gamma$. Clearly assuming a constant $\tilde{A}$ is a crude approximation for the actomyosin network, an approximation that will become less accurate close to the embryo poles. Nevertheless, our model captures the essence of the contraction process at the cellular scale and agrees well with the experimentally observed actomyosin dynamics.

During contraction, the density of a simple spring remains uniform along the spring's length. In modelling the cortical actomyosin network in this way, we therefore require that the density of contractile actomyosin is uniform across the contractile domain of length $l(t)$,

$$a(x,t) = \begin{cases} a_0 \dfrac{L}{l(t)} & 0 \leq x \leq l(t) \\ 0 & l(t) < x \leq L \end{cases}, \qquad (3)$$

where $a_0$ is the actomyosin density at $t=0$. Beyond the end of this domain we assume that there is no contractile actomyosin present, i.e. $a=0$. Initially, the contractile actomyosin occupies the entire cortex, i.e. $l(0)=L$. The position of the posterior end of the contractile actomyosin domain is calculated from Eq. 2, allowing $a(x,t)$ to be calculated from Eq. 3.

The presence of the anterior PAR proteins appears to enhance actomyosin contractility through an unknown mechanism (11). From Eq. 2 we see that this could take place through two effects. First, increased $A_m$ may allow the actomyosin network to contract to a shorter final length, acting to reduce $\lambda$. This effect is essential to achieve the different sizes of anterior domains that are seen in different mutants. Secondly, $A_m$ may act to change $\varepsilon$, altering the stiffness of the actomyosin network for a fixed natural length. In our model, the best agreement with experiment (with the exception of MEX-5/6 mutants, see below) is achieved when $\varepsilon$ remains constant, and where the effect of $A_m$ is to vary only the natural length, according to

$$\lambda(t) = \lambda_0 - \lambda_1 m(t), \qquad (4)$$

with $m(t)$ representing the contractile activity stimulated by the anterior PAR proteins. As discussed above, we take $m(t)$ to depend on the total amount of $A_m$ in the contractile region, given by

$$m(t) = \frac{1}{L}\int_0^{l(t)} A_m(x,t)\,dx. \qquad (5)$$

The assumption of linearity in Eq. 4 is not specifically required to reproduce the correct dynamics. With a suitable rescaling of $\lambda_1$ and the introduction of saturation of $m(t)$ (i.e. $m(t)$ tends to a constant when $A_m$ is large), quadratic or higher functions can be used with similar results.

With this model the magnitude of the local velocity at a given time, determined by the spring dynamics, is zero at the anterior pole and increases linearly towards the posterior until the end of the anterior actomyosin domain (see the Supplementary Information). The rate of contraction slows as a spring approaches its natural length, so the speed of the posterior end of the actomyosin region decreases over time. Both these properties appear consistent with experimental observations of the cortical actomyosin contraction pattern (11).

The similar and partially redundant CCCH finger proteins MEX-5 and MEX-6 are an important part of the signalling pathway that links PAR polarity to asymmetric gene expression (2). Surprisingly, the cytoplasmic MEX-5/6 proteins, which become polarized in response to PAR polarity, were also found to affect polarity establishment (9,10). Disrupting MEX-5/6 reduces the size and rate of expansion of the posterior PAR-2 domain. MEX-5/6 have been implicated in controlling protein degradation (23), and other finger motif proteins are thought to regulate RNA levels or translation rates (2,24-27). It is therefore possible that MEX-5/6 affect actomyosin dynamics indirectly by regulating the level of other factors that interact with the cytoskeleton. Consistent with this mechanism, the reduced rates of contraction in cells depleted of MEX-5/6 could be achieved in our actomyosin model by reducing the parameter $\varepsilon$ (data not shown).

Note that our simple model does not include actin polymerization or depolymerization reactions. While these processes may play a role in actomyosin reorganization, the defects observed in *nmy-2* depleted cells (9,28) suggest that the observed PAR dynamics is largely due to myosin-driven contraction. It is however possible that the actin turnover rate dictates the spontaneous dissociation rate of the anterior PAR proteins, (although it is thought that the anterior PAR proteins do not actually associate directly with the actin cytoskeleton). It appears unlikely that such a mechanism operates for the posterior PAR proteins, which are localized in regions of lower actin density.

**Wild-type dynamics**

Fig. 2 shows simulation results for the model described above as kymographs for the cortical density of actomyosin together with the cortical and cytoplasmic densities of the anterior and posterior PAR proteins. Initially, both anterior and posterior PAR proteins are present in the cytoplasm and at the cortex and are uniformly distributed along the cell length, as seen in experiment (9). Levels of $A_m$ and $P_c$ are slightly higher than $A_c$ and $P_m$ respectively. In our model, actomyosin contraction generates an anterior region where binding of the anterior PAR proteins is enhanced, and leaves a posterior region where cortical association of the anterior PAR proteins is greatly reduced. This eases the dissociation of the posterior PAR proteins at the posterior of the embryo, and hence the posterior PAR proteins become associated with the cortex at high levels here. The competition between the anterior and posterior PAR proteins means that each group excludes the other, thereby creating positive feedback allowing the density of whichever group is in the majority to increase. These reactions therefore give rise to the stably-polarized cortical distributions of the PAR proteins. Actomyosin contraction continues until ultimately the contractile domain is restricted to the anterior half of the embryo. Rapid initial contraction means that actomyosin quickly retracts to about 60% of the cell length within 3 to 4 minutes. The time to fully contract to mid-cell is approximately 8 minutes in our simulations, consistent with the time for which cortical and cytoplasmic flows are observed in vivo (10). The resulting cortical distributions show good agreement with experiment (9). The maximal velocity, at the posterior end of the contractile actomyosin region, is initially peaked at about 15$\mu m$ per minute, but rapidly drops to below 5$\mu m$ per minute. These speeds are comparable with reported flow speeds during contraction of 5-8$\mu m$ per minute (10,11,29).

**Mutant phenotypes**

Actomyosin dynamics and PAR localization in cells depleted of the different *par* proteins have previously been characterized experimentally (9-11). We have simulated the effects of the various mutants by making appropriate changes to the reaction scheme, discussed below. The results of these various changes are shown in Fig. 3.

In *par-3* mutants, PAR-6 and PKC-3 cannot associate with the cortex (9,15,16). In these cells, the posterior PAR proteins are uniformly distributed throughout the cortex (9,14), and actomyosin is cleared only from a small region around the posterior (11). We model this mutant by preventing the remaining anterior PAR proteins from associating with the cortex, setting $c_{A1}=c_{A2}=0$. This greatly suppresses actomyosin contraction, as shown in Fig. 3. Since the anterior PAR proteins cannot associate with the cortex, PAR-1 and PAR-2 are not excluded and hence accumulate uniformly at high levels, as seen in experiments. In our model, actomyosin contracts to approximately 85% of the embryo length, comparable to the experimentally measured actomyosin domain size of approximately 80% (11).

*par-6* and *pkc-3* mutants have similar phenotypes to *par-3* mutants (9,11). PAR-6 is required to localize PKC-3 to the cortex (15) and (according to our model) thereby stimulate cortical exclusion of PAR-1 and PAR-2. In the absence of PAR-6, PKC-3 remains in the cytoplasm while PAR-3 is seen to associate with the cortex at lower levels than in wild-type embryos (30). Similarly, in the absence of PKC-3, PAR-6 cannot become cortically localized (9,16), while PAR-3 is again weakly detected at the cortex (15,16). We assume that cortical association of the remaining anterior PAR proteins is disrupted in these mutants, possibly due to the loss of interaction between PAR-6 and CDC-42 (18,19). We modelled both *par-6* and *pkc-3* mutants by allowing A to associate with the cortex at a reduced rate, reducing $c_{A1}$ and $c_{A2}$ by a factor of 4. In addition, we prevent $A_m$ from excluding $P_m$, since cortical PKC-3 is required for this reaction. This was achieved by setting $c_{P4}=0$. We found that the model behaviour was then similar to the *par-3* simulations described above for the posterior PAR proteins and actomyosin (data not shown). The posterior PAR proteins are again uniformly distributed throughout the cortex, as observed experimentally for PAR-2 (9). Quantitative measurements of the extent of actomyosin contraction in these mutants have not been reported. The different localization patterns of PAR-3 and PAR-6/PKC-3 means that our assumption that the anterior PAR proteins function as a group is no longer valid. In implementing these mutants with the above changes we slightly underestimate the density of cytoplasmic PAR-6/PKC-3, since we assume that these proteins are removed from the cytoplasm when A associates with the cortex. However, in our model, PKC-3 only interacts with the posterior PAR proteins when cortically localized, while PAR-6 has no direct effect on the posterior PAR proteins. We can therefore simply interpret A as the density of PAR-3 in these mutant simulations.

In *par-1* mutants, the anterior PAR domain retracts beyond mid-cell (9). In our model, PAR-1 stimulates dissociation of the anterior PAR proteins. We simulate the *par-1* mutant by removing the competitive exclusion of $A_m$ by $P_m$, $c_{A4}=0$. PAR-2 is still able to associate with the cortex as in the wild-type (9,14), although in our model it cannot stimulate exclusion of $A_m$. According to our model, since the anterior PAR proteins are not actively excluded from the cortex, higher levels accumulate, which stimulates greater actomyosin contraction, as shown in Fig. 3. PAR-2 appears at the cortex at reduced levels relative to wild-type, due to faster exclusion by PKC-3. The actomyosin network and anterior PAR domain rapidly contract to mid-cell and ultimately occupy approximately the anterior 45% of the embryo. Our model therefore produces the correct qualitative change relative to the wild-type dynamics for the anterior PAR domain, although the size of this domain is slightly larger in our model than is observed experimentally (9). The extent of the actomyosin network in *par-1* mutants has not been reported. The initial rapid contraction of the anterior PAR domain appears somewhat faster than observed experimentally, where contraction beyond mid-cell takes approximately 6 minutes (9).

In *par-5* mutants the anterior and posterior PAR domains are seen to overlap (9,21). We assume that PAR-5 interacts with phosphorylated cortically-localized proteins and causes their dissociation. We therefore model this mutant by removing the competitive dissociation reactions between the cortical proteins, setting $c_{A4}=0$ and $c_{P4}=0$. This reproduces the overlapping domains of anterior and posterior PAR proteins observed experimentally, as shown in Fig. 3. The posterior PAR proteins remain uniformly localized, while the anterior

PAR proteins become mostly restricted to an anterior cortical domain. These observations appear consistent with experimental data (9), although the anterior PAR asymmetry appears somewhat more pronounced in our model than in experiments. In our simulations, *par-5* mutants show similar actomyosin contraction to *par-1* mutants. We are not aware of experimental measurements of the extent of actomyosin contraction in *par-5* mutants. Quantitative measurements of the PAR dynamics in *par-5* mutants are also complicated by the fact that the morphology of the cortex is much more irregular than in wild-type embryos (9).

Experiments in *par-2* mutants have provided evidence that actomyosin contraction is slightly reduced relative to wild-type, although not as dramatically as in anterior PAR protein mutants (11). Experimental measurements of the anterior PAR-6 domain in *par-2* mutants range from 50% (11) to 63% (9) of the cell length. PAR-2 has been suggested to promote cortical association of PAR-1 (14). We model this by reducing the cortical association rate of $P$, $c_{P1}$, by a factor of 3. However, this effect alone is not sufficient to reproduce the observed dynamics. The reduced association rate of $P$ leads to reduced cortical exclusion of $A_m$, and hence the anterior domain contracts beyond mid-cell in a similar way to the *par-1* mutant. This is qualitatively different from the reduced actomyosin contraction and expanded anterior PAR domain that are observed experimentally. Better agreement with the experimental dynamics can be achieved if, in addition to the reduced binding of PAR-1, we assume that PAR-1 is now more effective at excluding the anterior complex from the cortex than in the wild type. We included this effect by increasing the parameter $c_{A4}$ by a factor of 4. Now even though PAR-1 is present at the cortex at lower levels, it is still able to effectively reduce the amount of $A_m$ present. This result is shown in Fig. 3, where the anterior actomyosin and PAR domain both occupy approximately 60% of the embryo. The size of the anterior PAR domain is therefore comparable to experimental measurements (9,11).

In summary, our model gives generally good agreement with the experimentally observed mutant phenotypes for the cortical PAR protein distributions. This agreement is especially encouraging given the great simplicity of the model.

**Cytoplasmic polarity**

A key feature of development in the early *C. elegans* embryo is the polarization of cytoplasmic protein distributions, which leads to the asymmetric segregation of cytoplasmic proteins between daughter cells. The different cytoplasmic composition of these daughter cells leads to differentiation in development and cell fate. At the one-cell stage P-granules are restricted to the posterior, where they subsequently mark germline precursor cells (1). Moreover, as the cortical PAR domains form, MEX-5/6 become restricted to the anterior cytoplasm (2,9). The cytoplasmic distribution of the posterior PAR proteins also appears polarized, with a higher density at the posterior (9). PAR-1 has been suggested to negatively regulate MEX-5/6 activity, consistent with these proteins having oppositely polarized distributions (9). It is therefore important to test whether our model is able to account for this cytoplasmic polarity.

We added an additional equation to the model to describe the cytoplasmic density of MEX-5/6, $M$, as follows:

$$\frac{\partial M}{\partial t} = D_c \frac{\partial^2 M}{\partial x^2} + c_{M1} - c_{M2}M - c_{M3}MP_c \ . \qquad (6)$$

We assume that MEX-5/6 are uniformly produced at rate $c_{M1}$ and degraded spontaneously at rate $c_{M2}$. We also allow MEX-5/6 to be degraded by $P_c$ through the $c_{M3}MP_c$ term, consistent with negative regulation by PAR-1 (9). Since there is no experimental evidence for significant cortical levels of MEX-5/6, we restrict MEX-5/6 to interactions with cytoplasmic PAR-1. Kymographs of the cytoplasmic protein densities resulting from the model Eqs. 1a-d and 6 are shown in Fig. 2. As actomyosin contracts towards the anterior, the cytoplasmic distribution of the anterior PAR proteins also becomes polarized, with higher densities in the posterior cytoplasm. The posterior PAR proteins and MEX-5/6 are largely uniformly distributed, but with a slight increase in $P_c$ at the anterior and $M$ at the posterior. The cytoplasmic PAR distributions therefore have the opposite polarity to the cortical distributions. Hence, in our model, the cytoplasmic PAR-1, PAR-2 and MEX-5/6 polarities are the opposite of those observed experimentally. The model also produces a polarized cytoplasmic distribution of the anterior PAR proteins, whereas experimentally the cytoplasmic PAR-6 density appears uniform (9).

This behaviour is a result of the model structure and cannot be rectified by simply changing values of the model parameters. The anterior PAR proteins bind preferentially in the anterior, causing depletion of $A_c$ in the anterior relative to the posterior of the embryo. Dissociation of $A_m$ is also faster in the posterior than in the anterior due to exclusion by $P_m$, which tends to further increase levels of $A_c$ in the posterior part of the embryo. Similarly, dissociation of $P_m$ is faster in the anterior of the embryo, where levels of $A_m$ are high, than in the posterior. This leads to higher levels of $P_c$ in the anterior. We conclude that the simple model considered thus far cannot explain the observed cytoplasmic distributions of the PAR proteins. We therefore sought modifications to the model which gave better agreement with the experimental observations.

In the model described by Eqs. 1a-d, actomyosin contraction was coupled to PAR localization indirectly, through the density of actomyosin. However, actomyosin dynamics also drives large-scale cortical and cytoplasmic flows which affect the localization of the PAR proteins (11) and of cytoplasmic granules and vesicles (10,29). It is possible that these flows contribute to cytoplasmic polarity by localising the posterior PAR proteins to the posterior of the embryo. We tested the effects of these flows by introducing advection to the dynamic equations in addition to the reaction and diffusion terms described previously, to represent the directed motion of proteins. Further details of these changes are described in the Supplementary Information. The addition of these flows leads to minor changes in the transient PAR protein distributions during the initial period of rapid contraction. However, the steady-state distributions at the end of the contraction period were unchanged. We therefore conclude that these flows are unlikely to be important in establishing the correct cytoplasmic polarity.

The incorrect cytoplasmic polarity of the basic model appears in part because rapid competitive exclusion of cortical proteins increases the cytoplasmic density in the wrong half of the embryo. To overcome this effect we modified our model by introducing competitive degradation of the two PAR groups, perhaps due to the known phosphorylation reactions. Further details of the modified model can be found in the Supplementary Information. Such a model is able to give good agreement with all experimentally observed wild-type and mutant phenotypes (see Supplementary Figs. 3 and 4). However, in order to generate the observed polarized distributions the PAR proteins would have to be rapidly turned over, with a typical lifetime shorter than the actomyosin contraction timescale of a few minutes. Such a state would be extremely energetically expensive to maintain. For this reason, we believe that this mechanism is unlikely to be the correct explanation for the observed cytoplasmic PAR protein polarity.

*Cytoplasmic cytoskeletal asymmetry*
The polarization of the embryo cortex is driven by rearrangement of the cortical actomyosin network. It is therefore possible that the generation of cytoplasmic polarity is similarly driven by cytoskeletal rearrangement. PAR-2 is able to localize to the pronuclei or spindle and has been suggested to interact with microtubules (9,31). During the period of PAR polarity establishment, microtubules form primarily in the posterior part of the embryo as the pronuclei migrate and meet in the posterior (9,19). If the posterior PAR proteins are colocalized with the microtubules, this could effectively confine these proteins to the posterior cytoplasm. There is also evidence that cytoplasmic actin becomes largely restricted to the anterior (17). If the anterior PAR proteins are colocalized with the cytoplasmic actin, through a similar mechanism to their preferential localization to the anterior cortex, this mechanism could help to confine the cytoplasmic anterior PAR proteins to the anterior cytoplasm. Hence, this effect could neutralize the posterior polarity for $A_c$ found in our earlier model, and thus lead to a uniform distribution for $A_c$, as observed experimentally.

To test this mechanism, we modified the basic model in Eqs. 1a-d to introduce a second cytoplasmic state for the anterior and posterior PAR groups, $A_i$ and $P_i$ respectively. These variables represent proteins associated with the cytoplasmic cytoskeleton which are partly immobilized and also unable to bind to the cortex. We assumed that the local cytoplasmic actin density consists of two contributions, a constant component which is uniformly distributed throughout the embryo, and a varying component which moves with the cortical actomyosin network and has density proportional to $a(x,t)$. We therefore took the local cytoplasmic actin density to be proportional to $(1+c_a a(x,t))$. As a simple estimate, we assumed that the microtubule density is inversely related to the density of actomyosin, with the form $(1+c_a a(x,t))^{-1}$. However, our results are not specific to these particular choices for the cytoskeletal densities. We allowed anterior and posterior cytoplasmic PAR proteins to associate with the appropriate cytoplasmic cytoskeletal constituent at a rate proportional to the effective cytoskeletal density. We also assumed that the $A_i$ and $P_i$ were partly immobilized and could only diffuse slowly with the same diffusion constant $D_m$ as for the cortical proteins. The resulting equations are

$$\frac{\partial A_m}{\partial t} = D_m \frac{\partial^2 A_m}{\partial x^2} + (c_{A1} + c_{A2}a)A_c - c_{A3}A_m - c_{A4}A_m P_m \tag{7a}$$

$$\frac{\partial A_c}{\partial t} = D_c \frac{\partial^2 A_c}{\partial x^2} - (c_{A1} + c_{A2}a)A_c + c_{A3}A_m + c_{A4}A_m P_m - c_{A5}(1+c_a a)A_c + c_{A6}A_i \tag{7b}$$

$$\frac{\partial A_i}{\partial t} = D_m \frac{\partial^2 A_i}{\partial x^2} + c_{A5}(1+c_a a)A_c - c_{A6}A_i \tag{7c}$$

$$\frac{\partial P_m}{\partial t} = D_m \frac{\partial^2 P_m}{\partial x^2} + c_{P1}P_c - c_{P3}P_m - c_{P4}A_m P_m \tag{7d}$$

$$\frac{\partial P_c}{\partial t} = D_c \frac{\partial^2 P_c}{\partial x^2} - c_{P1}P_c + c_{P3}P_m + c_{P4}A_m P_m - \frac{c_{P5}}{1+c_a a}P_c + c_{P6}P_i \tag{7e}$$

$$\frac{\partial P_i}{\partial t} = D_m \frac{\partial^2 P_i}{\partial x^2} + \frac{c_{P5}}{1+c_a a}P_c - c_{P6}P_i \tag{7f}$$

$$\frac{\partial M}{\partial t} = D_c \frac{\partial^2 M}{\partial x^2} + c_{M1} - c_{M2}M - c_{M3}M(P_c + P_i). \tag{7g}$$

Fig. 4 confirms that such a mechanism is able to suitably polarize the distributions of cytoplasmic $P$ and MEX-5/6 and to generate a uniform cytoplasmic distribution of $A$, whilst retaining the cortical polarity of the basic model. To estimate the cytoplasmic cytoskeletal asymmetry required to generate the correct cytoplasmic distributions, we simulated Eqs. 7a-g and varied the asymmetry parameter $c_a$ (data not shown). To achieve the correct polarity for the cytoplasmic $P$ distribution, an anterior-posterior asymmetry in the density of microtubules of approximately a factor of two ($c_a=0.5\mu m$) was sufficient. The asymmetry of actin required to generate a uniform distribution of $A$ is somewhat larger at approximately a four-fold difference, because the incorrect polarization of $A_c$ in our initial model is more pronounced. The effectiveness of this mechanism is also dependent on the binding and dissociation kinetics, and achieving the correct polarity requires at least a certain fraction of the cytoplasmic proteins be immobilized. For a two-fold microtubule asymmetry, simulations with different binding ($c_{P5}$) and dissociation ($c_{P6}$) rates showed that at least a quarter of $P$ proteins in the posterior of the embryo must be in the immobilized form.

Simulations of the *par* mutants were also performed with this model, as described previously. In all cases, the behaviour of this model was essentially the same as the simple model of Eqs. 1a-d and 6 (data not shown). Hence, our new model is in good agreement with all the available experimental data on PAR polarization.

**DISCUSSION**

We have presented a mathematical model that couples interactions between the PAR proteins to actomyosin contraction, and largely reproduces the observed phenomenology of the PAR system at the one-cell stage of the *C. elegans* embryo. The cortical protein distributions in the wild-type and in *par*-depletion mutants can be explained through the experimentally reported interactions and with a mutual exclusion mechanism for the cortical PAR proteins proposed previously (9). Our modelling also confirms that polarization of the cortical actomyosin network is crucial for the correct establishment of polarity, restricting PAR-3/PAR-6/PKC-3 localization to the anterior, which in turn leads to polarization of PAR-1 and PAR-2 proteins. However, reproducing the correct cytoplasmic polarity of the PAR proteins is not straightforward. This issue has received surprisingly little attention, and the processes by which cytoplasmic polarity is generated are not understood. Regulating cytoplasmic polarity through MEX-5/6 and other CCCH-finger proteins is a vital function of the PAR system, crucial for the correct development of the different daughter cells. Our modelling clearly shows that the establishment of the correct cortical polarity is not sufficient to guarantee the appropriate cytoplasmic polarity of PAR-1/PAR-2 and MEX-5/6. We have therefore used modelling to quantitatively test additional mechanisms that could be involved in the generation of the correct cytoplasmic polarity.

We predict that asymmetry of the cytoskeleton in the cytoplasm drives the establishment of cytoplasmic protein polarity in parallel to the establishment of cortical PAR polarity by cortical cytoskeletal asymmetry. In this model, cytoplasmic actin becomes polarized in a similar way to the cortical actomyosin network, and retains the anterior PAR proteins in the anterior cytoplasm. At the same time, microtubules form primarily in the posterior and similarly localize the posterior PAR proteins to the posterior cytoplasm. These polarized cytoplasmic cytoskeletal distributions have previously been observed experimentally (9,17,19). This model is in agreement with the available data and makes a number of specific predictions. In particular, if microtubule polymerization could be disrupted the cytoplasmic polarity of the posterior PAR proteins should be reversed. Once again, this prediction should be directly testable since experiments to probe the role of microtubules would be possible without affecting cortical polarity. Our model also predicts that the observed uniform distribution of the anterior PAR proteins in the cytoplasm is in fact the result of a balance between two competing effects. The asymmetric binding and dissociation reactions included in our basic model tend to produce a posteriorly-polarized cytoplasmic distribution. However, binding to an anterior polarized distribution of cytoplasmic actin largely cancels this effect, leading to the uniform cytoplasmic distribution of the anterior PAR proteins that is observed experimentally. Testing this conclusion by disrupting the cytoplasmic actomyosin components without affecting cortical contraction would be difficult. However, it would be important to confirm the asymmetric cytoplasmic actin distribution suggested in (17). This mechanism can potentially explain cytoplasmic polarity during the pronuclear migration period, when the distribution of microtubules is biased towards the posterior of the embryo. However, it is not clear how polarity would be maintained after pronuclear meeting, when the distribution of microtubules becomes more uniform.

Questions also remain about how the polarized distributions of MEX-5/6 and other downstream proteins such as PIE-1 (2,9) are generated. An alternative mechanism for the generation of concentration gradients was recently suggested by Lipkow and Odde (32), whereby a protein is converted between two forms, which diffuse at different rates, by a localized activator and uniformly-distributed deactivator. This mechanism is able to generate a protein gradient opposite to that of the localized activator, as is typically seen for MEX-5/6 and PAR-1, if the activated protein is able to diffuse significantly faster than the unactivated form. If, instead of stimulating degradation, phosphorylation of MEX-5/6 by PAR-1 produces a phosphorylated form which is able to diffuse ~5 times faster than the unphosphorylated form, then this mechanism produces qualitatively similar MEX-5/6 gradients to those of our degradation mechanism in Eq. 6 (data not shown). Such a large change in the effective diffusivity suggests a significant change in the interactions of the protein, such as greatly reduced binding affinity for a sequestration reaction. Whether such a mechanism is actually important in *C. elegans* remains a question for future experiments.

The models discussed in this paper include a highly simplified description of the actomyosin network. While a detailed model of actomyosin activity may give a more mechanistic description of the contraction dynamics and smaller-scale phenomena such as cortical ruffling and pseudocleavage, we were able to capture the correct dynamics at the cellular scale important for cell polarity. The good agreement between the model and experiment supports the use of such a coarse-grained model, and shows that a more detailed model is not necessary to explain the polar organization of the PAR proteins. Our model does not, however, explain the secondary flows that are observed after pseudocleavage in *par* mutant embryos. In *par-2* mutants, actomyosin and the anterior PAR proteins flow back towards the posterior pole (9,11). In *par-1* and *par-5* mutants, the actomyosin distribution after pseudocleavage has not been reported, but the anterior PAR domain expands towards the posterior in both cases (9). It is therefore possible that our simple elastic model breaks down in this regime. A spring model in which the natural length is altered after pseudocleavage could potentially reproduce the correct PAR dynamics. However, it is not clear how the natural length in such a model should be determined. Munro et al (11) suggested that PAR-2 prevents re-expansion of the anterior domain after pseudocleavage by suppressing myosin binding. It is not clear why such a mechanism is not effective in *par-1* and *par-5* mutants, where PAR-2 is present at the cortex but posterior expansion of the anterior domain is observed. Alternatively, an inhomogeneous model including the posterior density of actomyosin together with myosin binding and unbinding reactions could potentially describe this behaviour.

It is not clear whether actomyosin contraction in the wild-type embryo specifically targets the mid-embryo position, whereby the boundary between the anterior and posterior domains scales with embryo length, as occurs, for example, in the *hunchback* expression boundary in the *Drosophila* embryo (33). Our model does not specifically self-organize to identify the mid-cell position – this must be achieved through appropriate parameter choices. However, scaling with embryo length can be achieved if the natural length in our actomyosin spring model is taken to be proportional to the embryo length. This can be achieved if the PAR protein and actomyosin densities remain constant as a function of embryo length. It would certainly be interesting to test the scaling properties of the anterior domain experimentally.

The model presented here deals specifically with the one-cell *C. elegans* embryo. One of the striking features of the PAR system is its conservation between different cell types and organisms (6-8). In many cases cell polarity and actin reorganization are linked (11,34,35), although we are not aware of any other examples where polarity establishment is accompanied by such large-scale rearrangement of cellular material. Our model suggests that these secondary cytoplasmic flows are not required to achieve the correct polarity, and that segregation of the actomyosin network together with competitive interactions between the PAR proteins are the keys to establishing PAR polarity. Some aspects of the model may therefore be directly applicable in other contexts.

## METHODS

### Simulations

Since in vivo concentrations of the PAR proteins are not known, concentrations are presented in arbitrary units of protein numbers per unit length, chosen such that the densities in the wild-type system are scaled to around $1\mu m^{-1}$. Simulations of Eqs. 1a-d and 6 were initialized with uniform concentrations $a=1\mu m^{-1}$, $A_c=0\mu m^{-1}$, $A_m=1\mu m^{-1}$, $P_c=1\mu m^{-1}$, $P_m=0\mu m^{-1}$, $M=1\mu m^{-1}$. The dynamic equations for the anterior and posterior PAR proteins and MEX-5/6 were integrated numerically on a lattice with spacing $\Delta x=0.2\mu m$ and with a fixed time interval of $\Delta t=10^{-3}s$. Smaller values were also tested and found not to alter the behaviour of the system, showing that any numerical instability was not significant. Simulations were run for 10 minutes with $v_l(t)$ set to zero, to allow the system to reach steady-state. This point is marked as $t=0$ in Figs. 2-4. The $t=0$ state in the wild-type simulations using Eqs. 1a-d and 6 is $A_c\approx0.4\mu m^{-1}$, $A_m\approx0.6\mu m^{-1}$, $P_c\approx0.6\mu m^{-1}$, $P_m\approx0.4\mu m^{-1}$, $M\approx1\mu m^{-1}$. The $t=0$ densities are different in the various mutant simulations, depending on the particular change to the dynamic equations. In each case there exists only one physical steady-state, so the choice of initial conditions is not significant.

Actomyosin contraction was initiated $t=0$. At each subsequent time step the contractile actomyosin activity, $m(t)$, and natural length, $\lambda(t)$, were calculated from Eqs. 5 and 4 respectively. These values were then used in Eqs. 2 and 3 to find $v_l(t)$ and the updated $l(t)$ and actomyosin density. The reaction and diffusion terms in Eqs. 1a-d and 6 were calculated with an explicit discretization scheme.

Parameter values were constrained to fit the dynamics observed in FRAP experiments (10). Otherwise, different parameter combinations were tested manually and selected by inspection to best match the wild-type and mutant behaviour. The qualitative model behaviour in wild-type simulations was robust to at least a 50% change in each reaction parameter individually. Parameters for the actomyosin network were selected to match the three cases of wild-type, *par-1* and *par-3* mutants. For the initial model in Figs. 2 and 3, using Eqs 1a-d and 6, the following parameter values were used: $L=50\mu m$, $a_0 = 1\mu m^{-1}$, $\lambda_0 = 42.5\mu m$, $\lambda_1 = 27\mu m^2$, $\varepsilon = 0.4\mu ms^{-1}$, $D_m = 0.25\mu m^2 s^{-1}$, $D_c = 5\mu m^2 s^{-1}$, $c_{A1} = 0.01s^{-1}$, $c_{A2} = 0.07\mu ms^{-1}$, $c_{A3} = 0.01s^{-1}$, $c_{A4} = 0.11\mu ms^{-1}$, $c_{P1} = 0.08s^{-1}$, $c_{P3} = 0.04s^{-1}$, $c_{P4} = 0.13\mu ms^{-1}$, $c_{M1} = 0.1\mu m^{-1}s^{-1}$, $c_{M2} = 0.02s^{-1}$, $c_{M3} = 0.135\mu ms^{-1}$. For the simulations shown in Fig. 4 for the model incorporating cytoplasmic immobilization, we used Eqs. 7a-g, where $c_a=5\mu m$, $c_{A1} = 0.013s^{-1}$, $c_{A2} = 0.091\mu ms^{-1}$, $c_{A5}=0.003s^{-1}$, $c_{A6}=0.06s^{-1}$, $c_{P1} = 0.096s^{-1}$, $c_{P5}=0.04s^{-1}$, $c_{P6}=0.04s^{-1}$, with the other parameters unchanged. The $t=0$ state in the wild-type simulations is $A_c\approx0.3\mu m^{-1}$, $A_m\approx0.6\mu m^{-1}$, $A_i\approx0.1\mu m^{-1}$, $P_c\approx0.5\mu m^{-1}$, $P_m\approx0.4\mu m^{-1}$, $P_i\approx0.1\mu m^{-1}$, $M\approx1\mu m^{-1}$.

ACKNOWLEDGEMENTS

F.T. is supported by the EPSRC. M.H. is supported by The Royal Society.

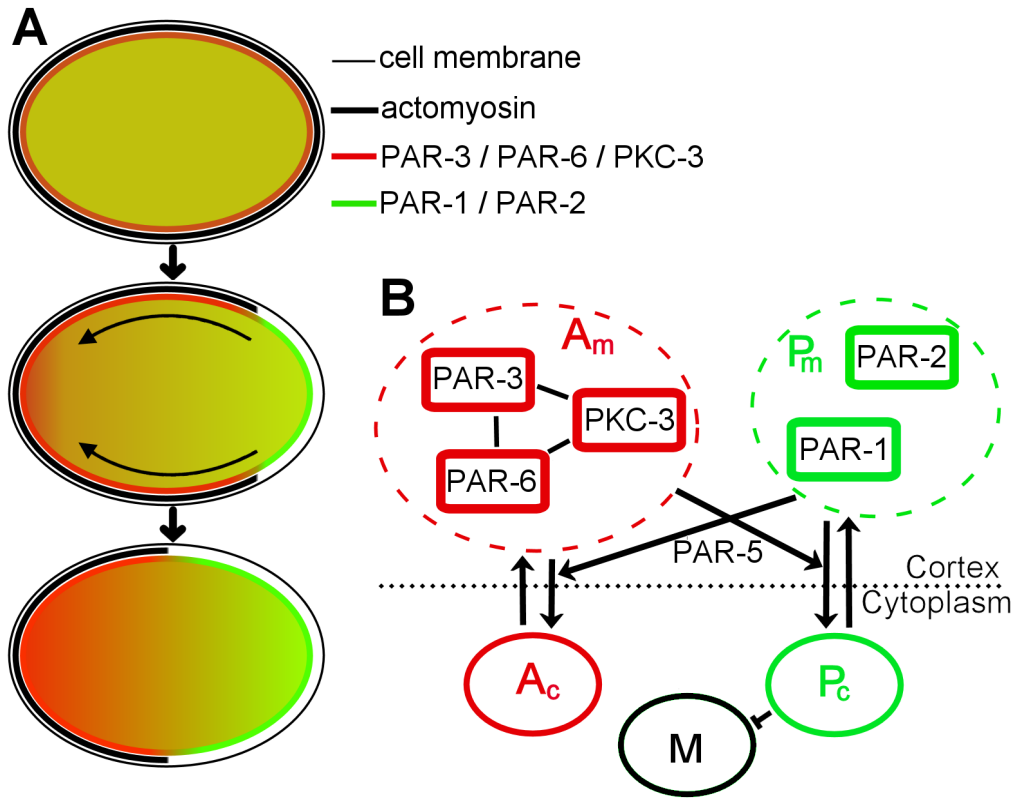

Figure 1: (*A*) Summary of PAR dynamics in wild-type embryos. Shown are the PAR distributions before, during, and after actomyosin contraction. Arrows indicate the direction of cortical actomyosin flow. The anterior pole is to the left. (*B*) Summary of the reaction scheme for the basic model in Eqs. 1a-d and 6. For clarity, actomyosin and the spatial aspects of the model are not shown.

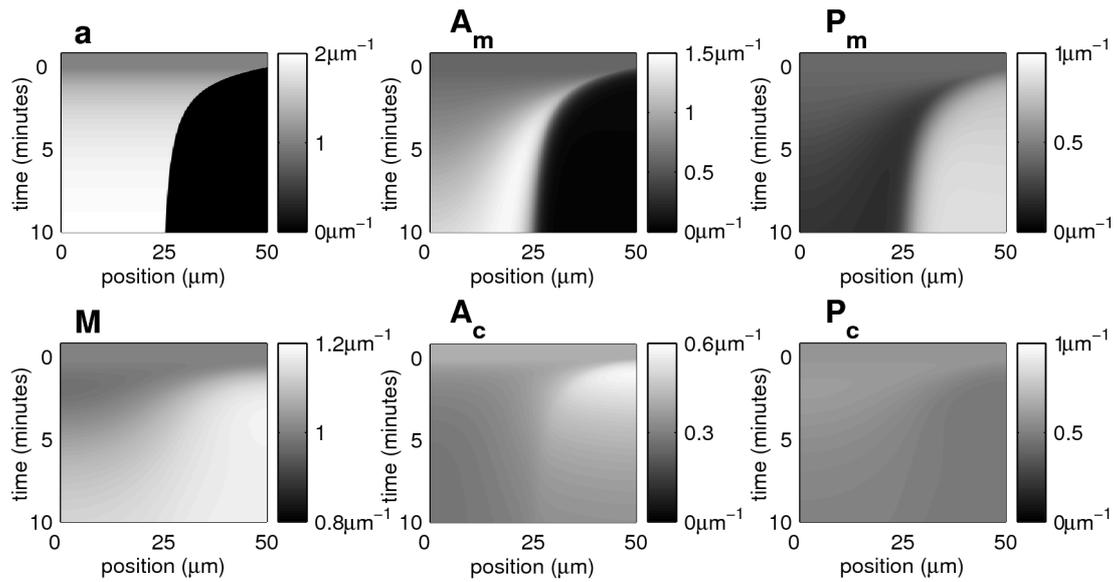

Figure 2: Wild-type simulation results for the model given by Eqs. 1a-d and 6. Kymographs showing the densities of: $a$, contractile actomyosin; $A_m$, cortically-localized anterior PAR proteins; $P_m$, cortically-localized posterior PAR proteins; $A_c$, anterior PAR proteins in the cytoplasm; $P_c$, posterior PAR proteins in the cytoplasm; and $M$, cytoplasmic MEX-5/6. The time marked as zero indicates the initiation time of actomyosin contraction. The greyscale is shown for each panel. Densities are presented in arbitrary units of $\mu m^{-1}$.

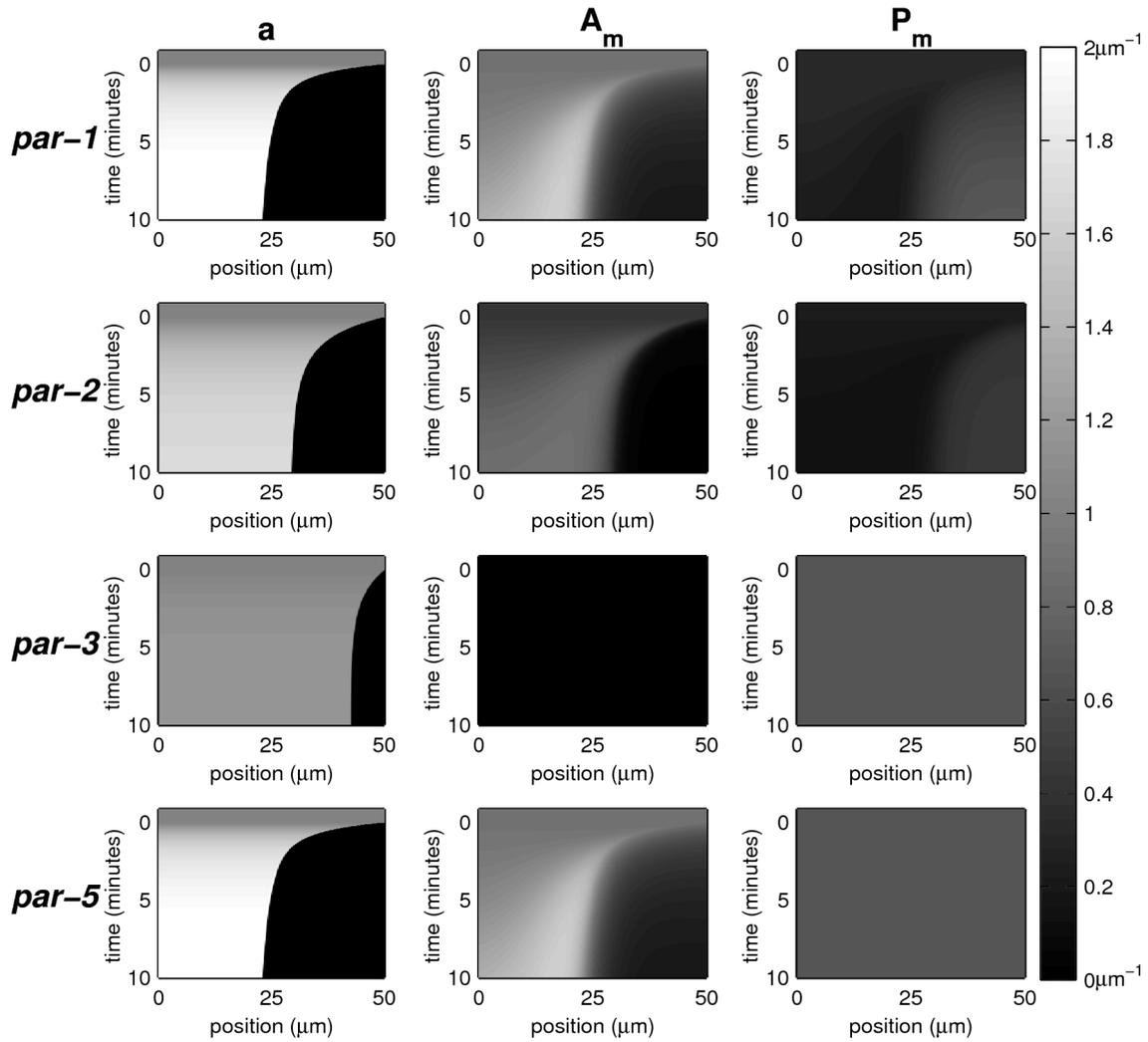

Figure 3: Cortical protein distributions in simulations of *par* mutant phenotypes. Simulations of Eqs. 1a-d and 6 were performed with modifications to represent depletion of the different *par* proteins, as described in the text. The greyscale indicated on the right was used for all panels.

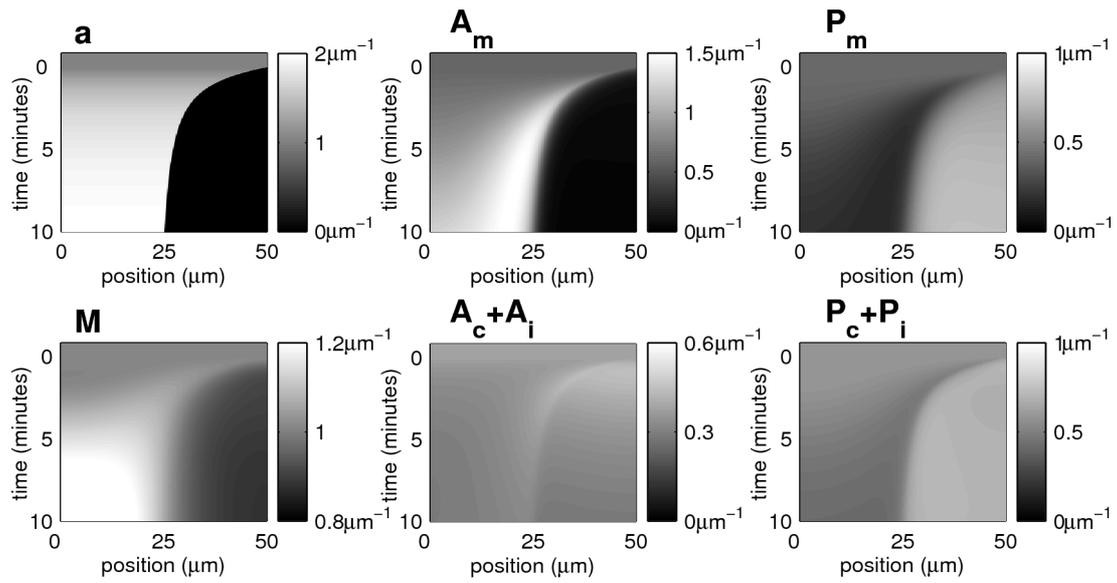

Figure 4: Simulation results for the model with partial immobilization of the cytoplasmic PAR proteins. $A_i$ and $P_i$ represent the densities of the partly immobilized cytoplasmic forms of the anterior and posterior PAR proteins respectively. In this case, approximately half of the cytoplasmic posterior PAR proteins were in the immobile form.

**Supplementary material for "Modelling the Establishment of PAR Protein Polarity in the One-Cell *C. elegans* Embryo"**

Filipe Tostevin & Martin Howard

**Cortical and cytoplasmic flows**

The motion of PAR proteins in cortical and cytoplasmic flows was modelled by adding advection to each of the model equations. These terms have the form $-\frac{\partial}{\partial x}(\rho_m v)$ for cortical proteins and $-\frac{\partial}{\partial x}(\rho_c v_c)$ for cytoplasmic proteins, where $\rho=A,P$ is the protein density and $v(x,t)$ and $v_c(x,t)$ are velocity fields for the cortical and cytoplasmic flows respectively. For example, Eq. 1a becomes

$$\frac{\partial A_m}{\partial t} = -\frac{\partial}{\partial x}(A_m v) + D_m \frac{\partial^2 A_m}{\partial x^2} + (c_{A1} + c_{A2} a) A_c - c_{A3} A_m - c_{A4} A_m P_m . \tag{S1}$$

The appropriate velocity field, *v*, in the contracting actomyosin region can be calculated directly from our actomyosin model. We consider the conservation equation for actomyosin,

$$\frac{\partial a}{\partial t} = -\frac{\partial}{\partial x}(av) . \tag{S2}$$

Since the density, *a*, remains uniform over $0 \leq x \leq l(t)$, $\frac{\partial a}{\partial t}$ must be the same everywhere in this region. This requires that $\frac{\partial v}{\partial x}$ also be uniform as a function of *x*. Finally, we can integrate and use the boundary conditions $v(0,t)=0$ and $v(l(t),t)=v_l(t)$ to find

$$v(x,t) = v_l(t)\left(\frac{x}{l(t)}\right) \quad 0 \leq x \leq l(t) , \tag{S3}$$

as we would expect for a uniform spring. The remaining cortical and cytoplasmic flows are not given by our actomyosin model, so we have simply assumed approximate forms for these flows. Other choices were also tested, and did not significantly alter the dynamics. We took the anterior-directed flow in the posterior cortex to be

$$v(x,t) = v_l(t)\left(\frac{L-x}{L-l(t)}\right) \quad l(t) < x \leq L . \tag{S4}$$

This ensures that *v* is continuous at $x=l(t)$, and that the flow speed goes to zero at the posterior pole. It is also in general agreement with experimental observations (1). We choose the cytoplasmic flow velocity in our model to be fastest near mid-cell, with the maximal flow velocity proportional to the maximal flow speed of the cortex:

$$v_c(x,t) = \begin{cases} -kv_l(t)\frac{x}{L/2} & 0 \leq x \leq L/2 \\ -kv_l(t)\left(\frac{L-x}{L/2}\right) & L/2 < x \leq L \end{cases} . \tag{S5}$$

These forms are broadly consistent with experimental observations (1), where it appears that cytoplasmic flow speeds are reduced near the poles. *k* is a parameter chosen to match the cytoplasmic velocity to that observed experimentally; we used $k=4/7$.

The advection of the PAR proteins was simulated with a first-order difference scheme, by calculating the flux between each pair of lattice sites. The change in density at each lattice site due to advection is given by

$$\Delta \rho_i = \begin{cases} -\dfrac{\Delta t}{\Delta x}(v_i \rho_i - v_{i-1}\rho_{i-1}) & if \quad v_i > 0 \\ -\dfrac{\Delta t}{\Delta x}(v_i \rho_{i+1} - v_{i-1}\rho_i) & if \quad v_i < 0 \end{cases}. \quad \text{(S6)}$$

Here $v_i$ represents the velocity at the boundary between sites $i$ and $i+1$. Boundary conditions were applied to ensure that $v(0,t)=v(L,t)=0$. Since the flows are relatively slow and smooth ($|(\Delta t)v_i| \ll \Delta x$), and unidirectional, we find that this discretization scheme remains well-behaved. A centred-difference scheme was also tested, with no change in the results.

The wild-type model dynamics with advection of the PAR proteins are shown in Supplementary Fig. 1. During the early part of the polarity establishment process, we can identify dynamic features in the PAR distributions which are the result of the advection of these proteins. As we would expect, cytoplasmic flows carry the cytoplasmic PAR proteins into the posterior, generating a transiently higher density of $A_c$ and $P_c$. Cortical flows also lead to a narrow, high-density, band of $P_m$ near the interface of the anterior/posterior PAR cortical domains. However, the stable polarized distributions that form at late times are unchanged. Assuming sufficiently fast cytoplasmic diffusion ($D_c$ larger than about $1\mu m^2 s^{-1}$), the system reaches a steady-state determined by diffusion and the protein interactions, whose timescales are short compared to the timescales over which cortical and cytoplasmic flows occur. We therefore conclude that movement of the PAR proteins in cortical and cytoplasmic flows likely cannot account for the observed cytoplasmic polarity in the embryo, and moreover, the flows lead to only minor transient changes in the cortical PAR distributions.

For completeness, simulations of the *par* mutants were also performed with cortical and cytoplasmic flows of the PAR proteins, as shown in Supplementary Fig. 2. The dynamics in *par-1* and *par-5* mutants in particular showed some transient differences during the early phase of the contraction dynamics. However, as in the wild-type simulations, the establishment of polarity and the final PAR distributions were unaffected.

**Competitive PAR protein degradation**

We added PAR protein production and degradation to the model as described in the main text. The model equations are now as follows:

$$\frac{\partial A_m}{\partial t} = D_m \frac{\partial^2 A_m}{\partial x^2} + (c_{A1} + c_{A2} a) A_c - c_{A3} A_m - c_{A4} A_m P_m \quad \text{(S7a)}$$

$$\frac{\partial A_c}{\partial t} = D_c \frac{\partial^2 A_c}{\partial x^2} - (c_{A1} + c_{A2} a + c_{A6}) A_c + c_{A3} A_m + c_{A5} - c_{A7} A_c P_c \quad \text{(S7b)}$$

$$\frac{\partial P_m}{\partial t} = D_m \frac{\partial^2 P_m}{\partial x^2} + c_{P1} P_c - c_{P3} P_m - c_{P4} A_m P_m \quad \text{(S7c)}$$

$$\frac{\partial P_c}{\partial t} = D_c \frac{\partial^2 P_c}{\partial x^2} - (c_{P1} + c_{P6}) P_c + c_{P3} P_m + c_{P5} - c_{P7} A_c P_c \quad \text{(S7d)}$$

$$\frac{\partial M}{\partial t} = D_c \frac{\partial^2 M}{\partial x^2} + c_{M1} - c_{M2} M - c_{M3} M P_c . \quad \text{(S7e)}$$

$c_{A5}$ is a constant production term for the anterior PAR proteins. Production of $P$ is similarly controlled by $c_{P5}$. $c_{A6}$ and $c_{P6}$ are spontaneous degradation rates for $A_c$ and $P_c$ respectively. $c_{A7} A_c P_c$ and $c_{P7} A_c P_c$ represent competitive degradation reactions between the anterior and posterior PAR proteins in the cytoplasm. Since we assume that cortical interactions lead to protein degradation, the $c_{A4}A_mP_m$ and $c_{P4}A_mP_m$ terms have been removed in Eqs. S7b and S7d respectively.

Since we have added production and degradation, the total protein levels can be altered by changing these parameters. In *par* mutants the degradation reactions can be disrupted, leading to different protein expression levels from wild-type. If levels of *A* increase significantly it would be possible for the natural length, $\lambda$ given by Eqs. 4 and 5, to become negative. Since this situation is unphysical, to remove this possibility we introduced saturation of $m(t)$ when levels of $A_m$ are high,

$$m(t) = \frac{\frac{1}{L}\int_0^{l(t)} A_m(x,t)dx}{1 + \sigma \frac{1}{L}\int_0^{l(t)} A_m(x,t)dx}. \tag{S8}$$

In simulations of this model we used the following parameters: $L=50\mu m$, $a_0 = 1\mu m^{-1}$, $\lambda_0 = 42.5\mu m$, $\lambda_1 = 60\mu m^2$, $\sigma = 1.75\mu m$, $\varepsilon = 0.35\mu ms^{-1}$, $D_m = 0.25\mu m^2 s^{-1}$, $D_c = 5\mu m^2 s^{-1}$, $c_{A1} = 0.008s^{-1}$, $c_{A2} = 0.072\mu ms^{-1}$, $c_{A3} = 0.032s^{-1}$, $c_{A4} = 0.008\mu ms^{-1}$, $c_{A5} = 0.1\ \mu m^{-1}s^{-1}$, $c_{A6} = 0.08s^{-1}$, $c_{A7} = 0.35\mu ms^{-1}$; $c_{P1} = 0.064s^{-1}$, $c_{P3} = 0.032s^{-1}$, $c_{P4} = 0.16\mu ms^{-1}$, $c_{P5} = 0.08\mu m^{-1}s^{-1}$, $c_{P6} = 0.06s^{-1}$, $c_{P7} = 0.016\mu ms^{-1}$, $c_{M1} = 0.1\mu m^{-1}s^{-1}$, $c_{M2} = 0.02s^{-1}$, $c_{M3} = 0.135\mu ms^{-1}$. With these parameters, the densities at $t=0$ are $A_c \approx 0.3\mu m^{-1}$, $A_m \approx 0.7\mu m^{-1}$, $P_c \approx 0.7\mu m^{-1}$, $P_m \approx 0.3\mu m^{-1}$, $M \approx 0.9\mu m^{-1}$. Simulation results for the wild-type are shown in Supplementary Fig. 3. The data show correctly polarised distributions of $P_c$ and $M$, and a slight anterior gradient of $A_c$.

Mutant simulations were also performed with this model, implemented as follows:
- *par-1*: We assume PAR-1 causes cortical and cytoplasmic degradation of the anterior PAR proteins and MEX-5/6. We therefore simulate this mutant by setting $c_{A4}=0$, $c_{A7}=0$, and $c_{M3}=0$.
- *par-2*: As for the basic model, we simulated this mutant by reducing the binding rate of PAR-1, $c_{P1}$, by a factor of 3 and increasing $c_{A4}$ by a factor of 4.
- *par-3*: As in the initial model, we prevent the anterior PAR proteins from associating with the cortex, $c_{A1}=0$ and $c_{A2}=0$.
- *par-5*: We assume PAR-5 is required for exclusion and degradation of cortical proteins. As in the initial model, we therefore simulated this mutant by setting $c_{A4}=0$ and $c_{P4}=0$. Cytoplasmic reactions were unchanged.

The results of mutant simulations are shown in Supplementary Fig. 4. In all cases, the extent of the anterior domain is consistent with experimental observations (2,3). The timescales for contraction are also consistent with experiment, except in the case of *par-1* for which contraction again appears slightly faster than observed experimentally (2).

Simulations of wild-type and mutants in this competitive degradation model were also performed with cortical and cytoplasmic flows included, as described above. The dynamics and PAR protein distributions were essentially unchanged from the results without flows (data not shown).

**Supplementary Figures**

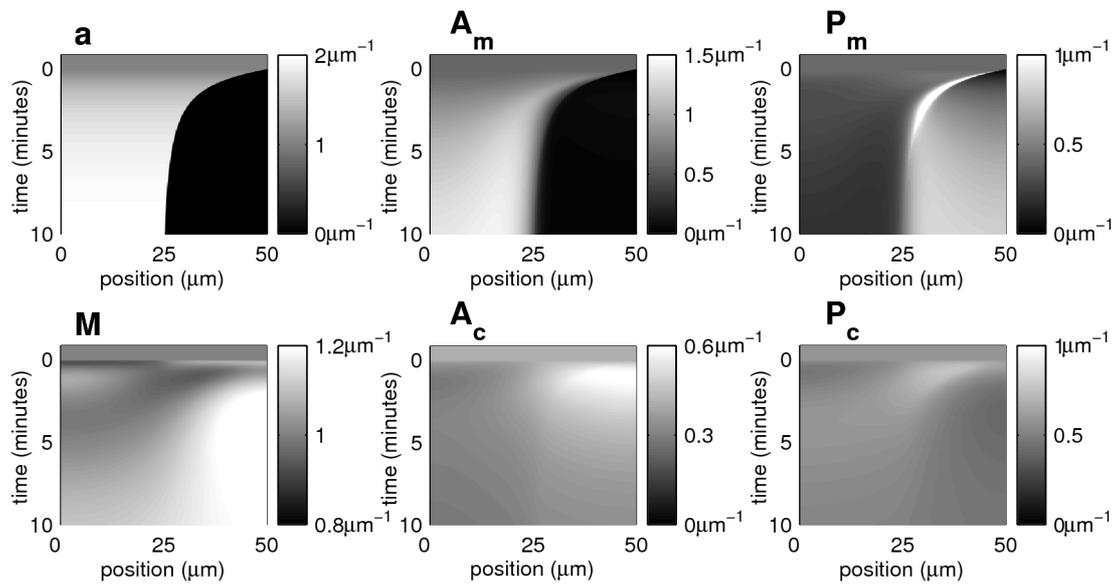

Supplementary Figure 1: Model dynamics with cortical and cytoplasmic flows. Simulations of the wild-type model were performed with advection terms added to Eqs. 1a-d and 6, as described in Eqs. S1-S6.

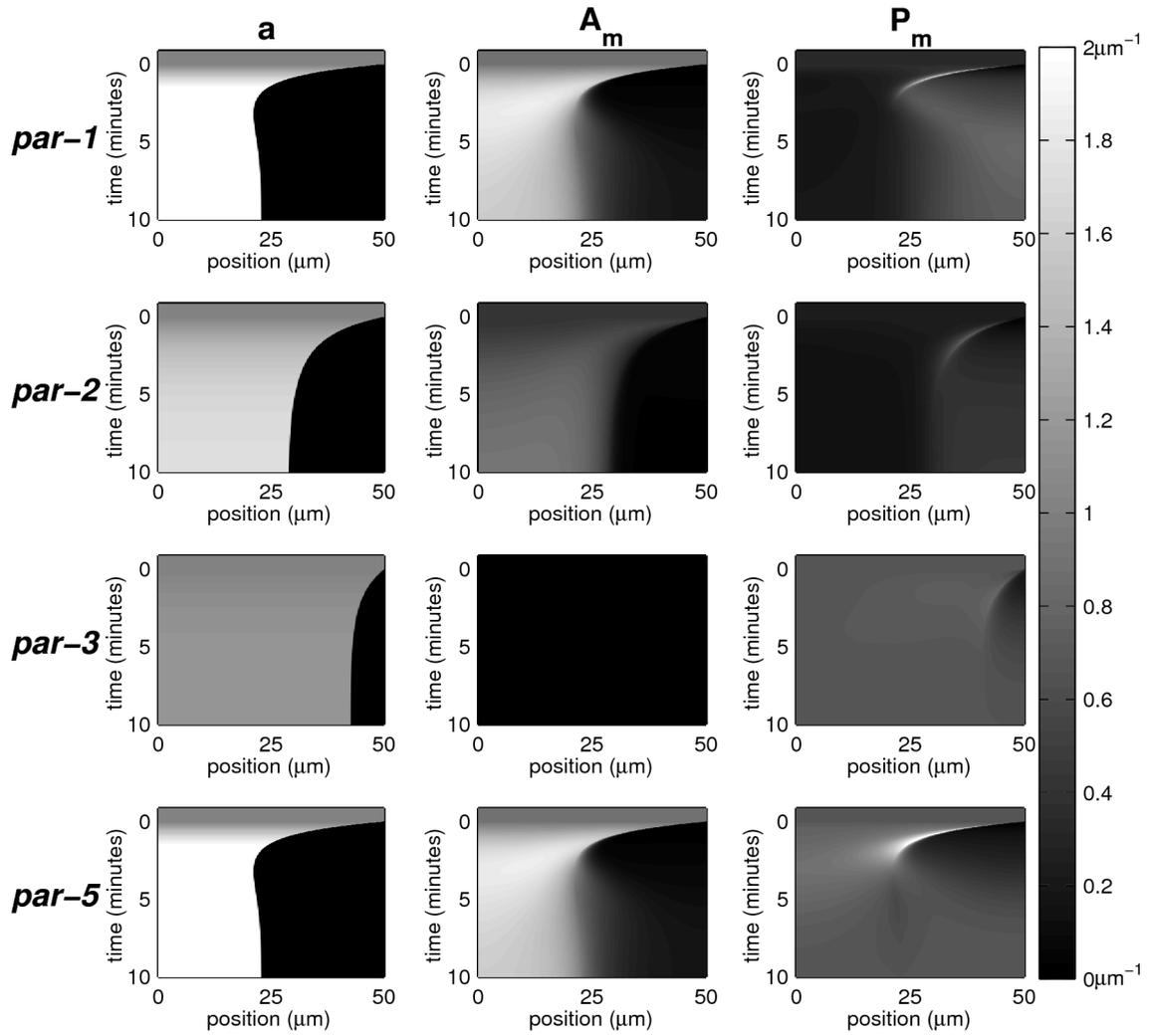

Supplementary Figure 2. Simulations of *par* mutants with cortical and cytoplasmic flows. The greyscale indicated on the right was used for all panels.

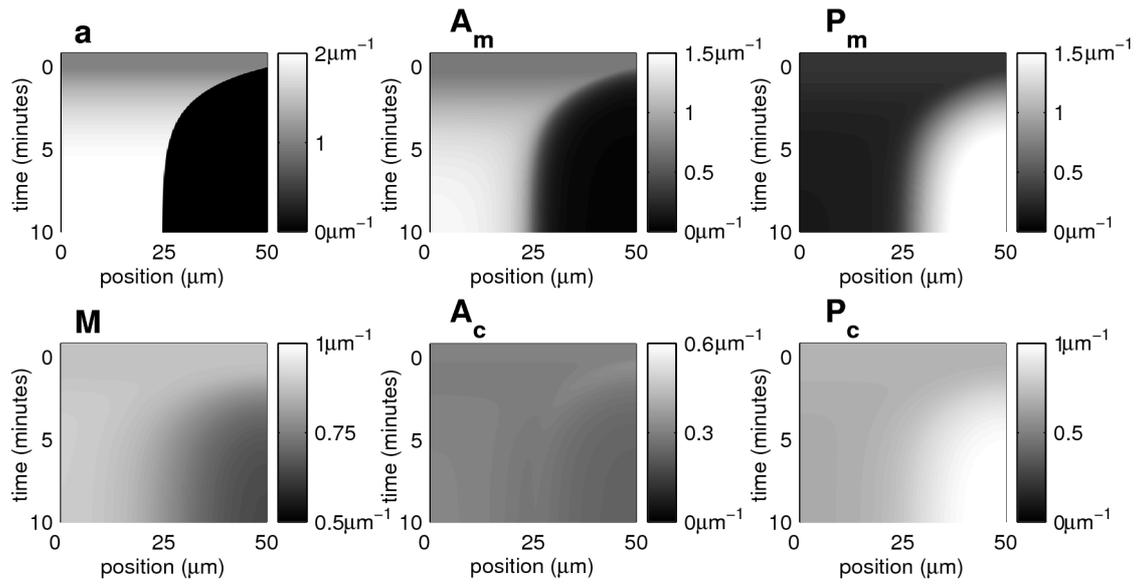

Supplementary Figure 3. Wild-type simulation results for the model with competitive protein degradation.

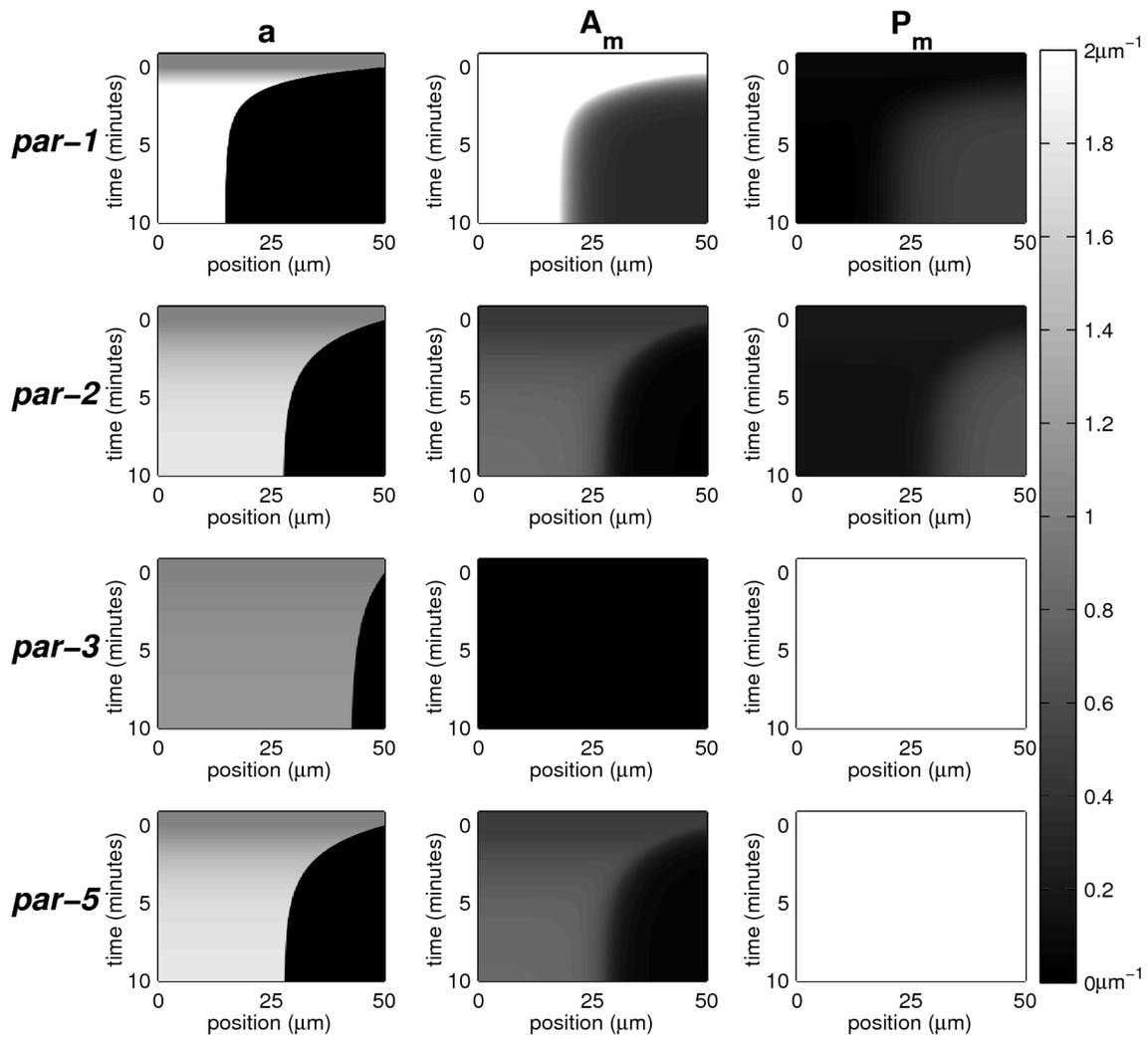

Supplementary Figure 4. Results for simulations of *par* mutants in the competitive degradation model. The greyscale indicated on the right was used for all panels.